\documentclass[%
aps,
prb,
reprint,
twocolumn
]{revtex4-2}

\usepackage[dvipdfmx]{graphicx}
\usepackage{mathtools}
\begin{document}


\title{Critical quantum states and hierarchical spectral statistics in a Cantor potential}


\author{F. Iwase}
\email[]{iwasef@tokyo-med.ac.jp}
\affiliation{Department of Physics, Tokyo Medical University\\
 6-1-1 Shinjuku, Shinjuku-ku, Tokyo 160-8402, Japan}


\date{\today}

\begin{abstract}
We study the spectral statistics and wave-function properties of a one-dimensional quantum system subject to a Cantor-type fractal potential.
By analyzing the nearest-neighbor level spacings, inverse participation ratio (IPR), and the scaling behavior of the integrated density of states (IDS), we demonstrate how the self-similar geometry of the potential is imprinted on the quantum spectrum.
The energy-resolved level spacings form a hierarchical, filamentary structure, in sharp contrast to those of periodic and random systems.
The normalized level-spacing distribution exhibits a bimodal structure, reflecting the deterministic recurrence of spectral gaps.
A multifractal analysis of eigenstates reveals critical behavior: the generalized fractal dimensions $D_q$ lie strictly between the limits of extended and localized states, exhibiting a distinct $q$-dependence.
Consistently, the IPR indicates the coexistence of quasi-extended and localized features, characteristic of critical wave functions.
The IDS shows anomalous power-law scaling at low energies, with an exponent close to the Hausdorff dimension of the underlying Cantor set, indicating that the geometric fractality governs the spectral dimensionality.
At higher energies, this scaling crosses over to the semiclassical Weyl law.
Our results establish a direct connection between deterministic fractal geometry, hierarchical spectral statistics, and quantum criticality.
\end{abstract}


\maketitle

\section{Introduction\label{sec:intro}}

Fractal geometry provides a ubiquitous framework for describing complex physical systems characterized by self-similarity, ranging from disordered solids and quasiperiodic lattices to photonic and acoustic metamaterials~\cite{Mandelbrot1982, Falconer2014}.
Within this landscape, the Cantor set stands as a paradigmatic example of a deterministic fractal~\cite{Falconer2014}.
Embedding such a fractal structure into a quantum potential creates a unique platform for exploring the intricate interplay between geometrical self-similarity and quantum interference, particularly regarding how the underlying geometry dictates spectral properties and wave-function statistics.

One-dimensional quantum systems serve as an ideal testbed for elucidating these fundamental issues.
In perfectly periodic potentials, Bloch's theorem guarantees the formation of continuous energy bands separated by well-defined gaps, resulting in extended eigenstates~\cite{Kittel2005}.
Conversely, in uncorrelated random potentials, the scaling theory of localization predicts that all eigenstates are exponentially localized, even for infinitesimal disorder strength (Anderson localization)~\cite{Anderson1958}.
Situated between these two extremes, quasiperiodic and fractal potentials host a rich variety of unconventional spectral features, such as singular continuous spectra and critical wave functions that are neither fully extended nor localized~\cite{Kohmoto1983, Ostlund1983, Evers2008}.
The Cantor potential is particularly significant in this context because it breaks translational symmetry entirely while strictly preserving long-range order through its recursive, deterministic construction.

While previous studies have extensively investigated the spectral and transport properties of fractal and quasiperiodic systems---revealing Cantor-like energy spectra and anomalous diffusion~\cite{Ostlund1983, Kohmoto1987, Ketzmerick1997}---a direct and unified comparison of these systems remains incomplete.
Specifically, although the hierarchical gap structures inherent to Cantor-type potentials have been mathematically characterized~\cite{Hofstadter1976, Ostlund1983, Kohmoto1987}, a systematic cross-examination against periodic and random benchmarks using modern numerical diagnostics is lacking.
In this paper, we bridge this gap by performing a comprehensive analysis of level statistics, inverse participation ratio (IPR), multifractal dimensions, and the scaling behavior of the integrated density of states (IDS).
Moreover, given the recent experimental advances in fabricating artificial potentials using photonic waveguides~\cite{Lahini2009,Biesenthal2022}, acoustic metamaterials~\cite{Xiao2015,Zhao2018}, and cold atoms~\cite{Roati2008}, our results provide practical guidelines for engineering quantum states with tunable localization properties.
By controlling the fractal dimension of the potential, one can deterministically design the spectral dimensionality and confinement strength of the system.
Our approach clarifies how the deterministic fractal geometry directly manifests in quantum critical statistics, distinguishing the Cantor system from both periodic order and random disorder.

In this work, we perform a comprehensive numerical study of a one-dimensional quantum particle subject to a Cantor potential and systematically compare the results with those obtained for periodic and random potentials constructed with comparable parameters.
By analyzing the energy spectra, density of states (DOS), level-spacing statistics, IPR, and multifractal dimensions, we clarify how the fractal geometry dictates the spectral fluctuations and wave-function localization properties.
Particular attention is paid to the scaling behavior of the IDS, which provides direct insight into the relationship between the spectral dimensionality and the fractal dimension of the underlying potential.

We demonstrate that the Cantor potential exhibits characteristic hierarchical, filamentary structures in the energy-gap distribution that are absent in random systems, reflecting its deterministic, self-similar nature.
The normalized level-spacing distribution displays a bimodal structure, indicating the coexistence of dense and sparse spectral regions arising from the fractal gap generation.
Furthermore, our multifractal analysis reveals that the eigenstates are neither fully extended nor exponentially localized but exhibit critical behavior with a spectrum of generalized fractal dimensions.
Crucially, the scaling of the IDS at low energies follows a power law with an exponent close to the Hausdorff dimension of the Cantor set, whereas at high energies it converges to the semiclassical Weyl law.
This observation indicates a crossover from fractal-dominated spectral dimensionality to classical behavior driven by short-wavelength kinetics.

This paper is organized as follows. 
Section \ref{sec:method} describes the model Hamiltonian and the numerical methods employed.
In Sec. \ref{sec:results}, we present the numerical results, focusing on spectral properties, gap statistics, multifractality, and scaling behavior of the IDS.
Section \ref{sec:discussion} discusses the physical implications of these findings, particularly in the context of spectral dimensionality and the distinction from disorder-induced localization.
Finally, Sec. \ref{sec:conclusion} summarizes our conclusions.

\section{Model and Method of Calculations\label{sec:method}}
\subsection{Hamiltonian and boundary conditions}

We consider a quantum particle of mass $m$ confined in a one-dimensional box of length $L$, subject to an external potential $V(x)$.
The system is governed by the time-independent Schr\"odinger equation:
\begin{equation}
-\frac{\hbar^2}{2m}\frac{d^2\psi(x)}{dx^2} + V(x)\psi(x) = E\psi(x)
\end{equation}
where $E$ is the energy eigenvalue and $\psi(x)$ is the corresponding eigenfunction.
We set $L=1$, and impose Dirichlet (hard-wall) boundary conditions at the edges of the domain:
\begin{equation}
\psi(0) = \psi(L) = 0.
\end{equation}
This setup models a particle in a finite potential well with internal structural modulation.

\subsection{Construction of potentials}
To elucidate the effects of fractal geometry on quantum states, we analyze three distinct potential profiles: a periodic potential, a random potential, and a Cantor fractal potential.
Crucially, all three potentials are constructed using the same fundamental building blocks---rectangular barriers of identical height $V_0$ and width $w$---and the same total number of barriers $N_\mathrm{b}$.
This ``equal footing'' approach ensures that any differences in spectral or localization properties arise solely from the spatial arrangement of the potential barriers (ordered, disordered, or fractal), rather from variations in the average potential strength.

\subsubsection{Periodic potential}
The periodic potential, $V_\mathrm{p}(x)$, consists of a regular array of $N_\mathrm{b}$ identical rectangular barriers placed at equal intervals within the domain $[0, L]$.
This configuration serves as a reference model for a perfectly ordered lattice.
By matching the barrier parameters ($V_0$, $w$, $N_\mathrm{b}$) to those of the Cantor potential (defined below), we ensure a fair comparison.
Such a periodic arrangement naturally leads to Bloch-type eigenstates and well-defined energy bands, providing a benchmark against which the deterministic aperiodic and disordered systems can be evaluated.

\subsubsection{Random potential}
The random potential, $V_\mathrm{r}(x)$, is composed of the same $N_\mathrm{b}$ barriers as the periodic case, but their positions are uncorrelated.
The centers of the barriers are drawn from a uniform probability distribution over the interval $[0, L]$, subject to a non-overlapping constraint (hard-core disorder).
Since the total number of barriers, their individual widths, and heights are identical to those in the periodic and Cantor models, this potential isolates the effects of pure spatial disorder.
This model represents an amorphous medium and is expected to exhibit Anderson localization \cite{Anderson1958, Mott1961}.

\subsubsection{Cantor potential}
The Cantor potential, $V_\mathrm{C}(x)$, is constructed iteratively following the standard middle-third Cantor set generation rule.
At generation $n=0$, a single rectangular potential barrier of height $V_0$ covers the entire interval $[0, L]$.
In each subsequent generation, the middle third of every existing barrier is removed (setting $V(x)=0$), leaving two segments of equal length.
Consequently, at generation $n$, the potential consists of $N_\mathrm{b} = 2^n$ barriers, each with a width $L/3^n$.
This construction yields a self-similar fractal structure with a Hausdorff dimension of $D_\mathrm{H} = \log 2/\log 3 \approx 0.631$.

In this study, we focus on the seventh generation ($n=7$), which corresponds to $N_\mathrm{b} = 128$ barriers.
The total fraction of the length occupied by the potential is $(2/3)^7 \approx 5.9\%$.
This generation level provides a sufficiently developed fractal hierarchy to observe scaling behavior while maintaining computational feasibility.
The corresponding periodic and random potentials are thus also constructed with $N_\mathrm{b} = 128$ barriers of width $w = L/3^7$.

\subsection{Numerical diagonalization}
In our numerical calculations, we employ a system of units where $\hbar = 1$ and $2m = 1$.
The spatial domain is fixed to $L=1$ and discretized into $N=20,000$ internal grid points with a uniform lattice spacing $\Delta x = L/(N+1)$.
The Hamiltonian is constructed using the standard second-order central finite-difference scheme for the kinetic energy operator.
This results in a sparse tridiagonal matrix representation of the Hamiltonian $H$, with matrix elements given by:
\begin{equation}
    H_{ij} =
    \begin{cases}
        \frac{2}{\Delta x^2} + V(x_i) & \text{if } i=j, \\
        -\frac{1}{\Delta x^2} & \text{if } |i-j|=1, \\
        0 & \text{otherwise},
    \end{cases}
\end{equation}
where $x_i=i\Delta x$ denotes the position of the $i$-th grid point.

The height of the potential barriers is defined as 
\begin{equation}
V_0 = \frac{c}{\Delta x^2},
\end{equation}
where $c$ is dimensionless parameter controlling the potential strength relative to the kinetic energy scale.
In this study, we set $c=0.1$.
This specific scaling $V_0 \propto \Delta x^{-2}$ is chosen to ensure that the ratio between the potential strength and the kinetic energy hopping amplitude (which scales as $1/\Delta x^2$) remains constant.
In this regime, the discretized system effectively maps onto a tight-binding model with on-site potentials and nearest-neighbor hopping, allowing us to investigate the structural effects of the fractal geometry independent of the discretization scale.
Although the discretization introduces an energy scale of order $10^5$ in the high-energy region, this is a natural consequence of the finite-difference representation and does not indicate any numerical instability.
The eigenvalues were obtained using standard sparse diagonalization methods.
We compute the lowest $K=2000$ eigenvalues and their corresponding eigenfunctions.
We have confirmed that the statistical properties of the spectra and wave functions are robust against changes in the grid size $N$.

\subsection{Analysis methods}
\subsubsection{Multifractality of wave functions}
To characterize the spatial structure and fractal nature of the eigenstates, we perform a multifractal analysis of the wave functions.
The concept of multifractality was originally introduced by Mandelbrot in the context of turbulence~\cite{Mandelbrot1974}.
This formalism was later generalized~\cite{Halsey1986} and has become a powerful tool to investigate critical phenomena in Anderson localization and quantum Hall systems~\cite{Evers2008}.

For a one-dimensional system of length $L$, the generalized inverse participation ratio (gIPR), denoted as $P_q$, is defined by the integral of the $q$-th moment of the probability density:
\begin{equation}
  P_q = \int_0^L |\psi_k(x)|^{2q} dx
\end{equation}
where $\psi_k(x)$ is the eigenfunction of the $k$-th state.
For a multifractal wave function, $P_q$ exhibits a power-law scaling with respect to the system resolution (or system size in lattice models).
In our discretized system with $N$ grid points, this scaling relation is expressed as:
\begin{equation}
  P_q \sim N^{-(q-1)D_q},\label{eq:scaling_Pq}
\end{equation}
where $D_q$ is the generalized fractal dimension.
In the thermodynamic limit ($N \to \infty$), $D_q \to 1$ corresponds to an extended state, while $D_q \to 0$ implies a localized state.
Values strictly in the range $0 < D_q < 1$ that persist upon scaling indicate a multifractal (critical) state.
However, in finite-size systems, calculated dimensions may exhibit effective intermediate values due to crossover or finite localization lengths.

For a finite system size $N$, the generalized dimension $D_q$ can be estimated via:
\begin{equation}
    D_q \simeq - \frac{\ln P_q}{(q-1)\ln N} \label{eq:Dq}
\end{equation}
Note that, for $q=2$, $P_2$ corresponds to the standard IPR.
A large IPR implies strong localization, whereas a value inversely proportional to the system size indicates delocalization.

In the numerical implementation, we evaluate the ensemble-averaged moments to suppress state-to-state fluctuations.
The discretized version of the $q$-th moment for the $k$-th eigenstate, $\tilde{P}_q^{(k)}$, is calculated as:
\begin{equation}
\tilde{P}_q^{(k)} = \sum_{i=1}^{N} |\psi_k(x_i)|^{2q} \Delta x
\end{equation}
We then compute the average over the lowest $K$ eigenstates:
\begin{equation}
\langle P_q \rangle= \frac{1}{K} \sum_{k=1}^{K} \tilde{P}_q^{(k)}
\end{equation}
This averaged quantity $\langle P_q \rangle$ is used to determine the generalized fractal dimensions $D_q$ by substituting $\langle P_q \rangle$ for $P_q$ in Eq.~(\ref{eq:Dq}).

\subsubsection{Exclusion of boundary-induced states}
In finite systems subject to Dirichlet (hard-wall) boundary conditions, eigenstates localized near the system edges frequently emerge, particularly in the periodic and random potential models.
These boundary-induced states are artifacts of the confinement geometry and do not reflect the intrinsic bulk properties of the underlying potential structures.
Since the primary objective of this study is to elucidate the bulk spectral statistics and multifractality, we systematically identify and exclude these states from our analysis.

Specifically, we classify an eigenstate as a boundary artifact if it satisfies both of the following criteria:
\begin{enumerate}
    \item[(i)] Probability concentration: More than 20\% of the total probability density is localized within 200 grid points (corresponding to 1\% of the system length $L$) from either boundary.
    \item[(ii)] Center of mass: The center of mass of the probability distribution, defined as $x_\mathrm{cm} = \int_0^L x |\psi_k(x)|^2 dx$, lies within the regions $x_\mathrm{cm} \le 0.1L$ or $x_\mathrm{cm} \ge 0.9L$.
\end{enumerate}
These thresholds were chosen to reliably filter out states pinned to the walls while retaining bulk-localized states.
We have confirmed that the main conclusions presented in this work are robust against moderate variations in these threshold parameters.

\subsubsection{Integrated density of states}
The IDS, denoted here as $\mathcal{N}(E)$, is defined as the normalized cumulative count of eigenvalues less than or equal to a given energy $E$.
It is written as 
\begin{equation}
\mathcal{N}(E) = \frac{1}{N}\sum_n \Theta(E-E_n),
\end{equation}
where $N$ is the total number of eigenstates (equal to the number of grid points) and $\Theta(x)$ is the Heaviside step function.
In this study, we evaluate $\mathcal{N}(E)$ numerically using the discrete spectrum obtained from the finite-size diagonalization.

For systems with self-similar potentials, the energy spectrum is often singular continuous, and the IDS exhibits a characteristic ``Devil's staircase'' structure~\cite{Ostlund1983}.
In the vicinity of the spectral edges (low energy edges), the IDS is expected to follow a power-law scaling:
\begin{equation}
\mathcal{N}(E) \sim (E - E_\mathrm{min})^\alpha,
\end{equation}
where $E_\mathrm{min}$ is the ground-state energy.
The exponent $\alpha$ is a critical quantity that reflects the effective spectral dimension and the fractal geometry of the underlying potential.

We pay particular attention to the scaling behavior in the low-energy regime.
In this analysis, the energy is measured relative to the ground-state eigenvalue $E_\mathrm{min}$.
This referencing is essential because, in a finite system under Dirichlet boundary conditions, the ground-state energy is non-zero (zero-point energy).
Subtracting this shift allows for a proper characterization of the intrinsic scaling properties of the spectrum near its onset.

\subsubsection{Level-spacing analysis}
To analyze the spectral fluctuations and correlations, we perform two complementary types of gap analysis.
First, we examine the energy dependence of the nearest-neighbor level spacings, defined as:
\begin{equation}
\Delta E_i = E_{i+1} - E_i.
\end{equation}
This quantity directly reflects the local fine structure of the energy spectrum and reveals the hierarchical gap structure characteristic of fractal potentials.

Second, we investigate the distribution of the normalized level spacings, $s_i$, to compare the results with universal predictions from random matrix theory (RMT)~\cite{Mehta2004, Bohigas1984}. 
This requires an ``unfolding'' procedure to remove the effects of the global variation in the average DOS, thereby isolating the intrinsic local fluctuations~\cite{Guhr1998, Gomez2002}.
The normalized spacing is defined as:
\begin{equation}
s_i = \frac{\Delta E_i}{\langle \Delta E \rangle},
\end{equation}
where $\langle \Delta E \rangle$ represents the local mean spacing in the vicinity of energy $E_i$.
By construction, the mean value of the normalized spacings is unity, i.e., $\langle s_i \rangle = 1$.

In our numerical implementation, the local mean spacing $\langle \Delta E \rangle$ is estimated from the local DOS, defined as the slope of the smoothed IDS, $\bar{\mathcal{N}}(E)$.
We obtained $\bar{\mathcal{N}}(E)$ by fitting the numerical IDS with a cubic smoothing spline.
We confirmed that the resulting spacing distribution is insensitive to moderate variations in the smoothing parameters.
Values of $s_i > 1$ indicate spectral regions that are locally sparse (level repulsion or gaps), whereas $s_i < 1$ correspond to locally dense spectral regions (level clustering).
This normalized gap analysis provides a quantitative measure of how the fractal geometry modulates the statistical properties of the quantum spectrum.

\section{Results\label{sec:results}}
\subsection{Potential profiles}

\begin{figure*}[t]
\begin{center}
\includegraphics[width=17.5cm]{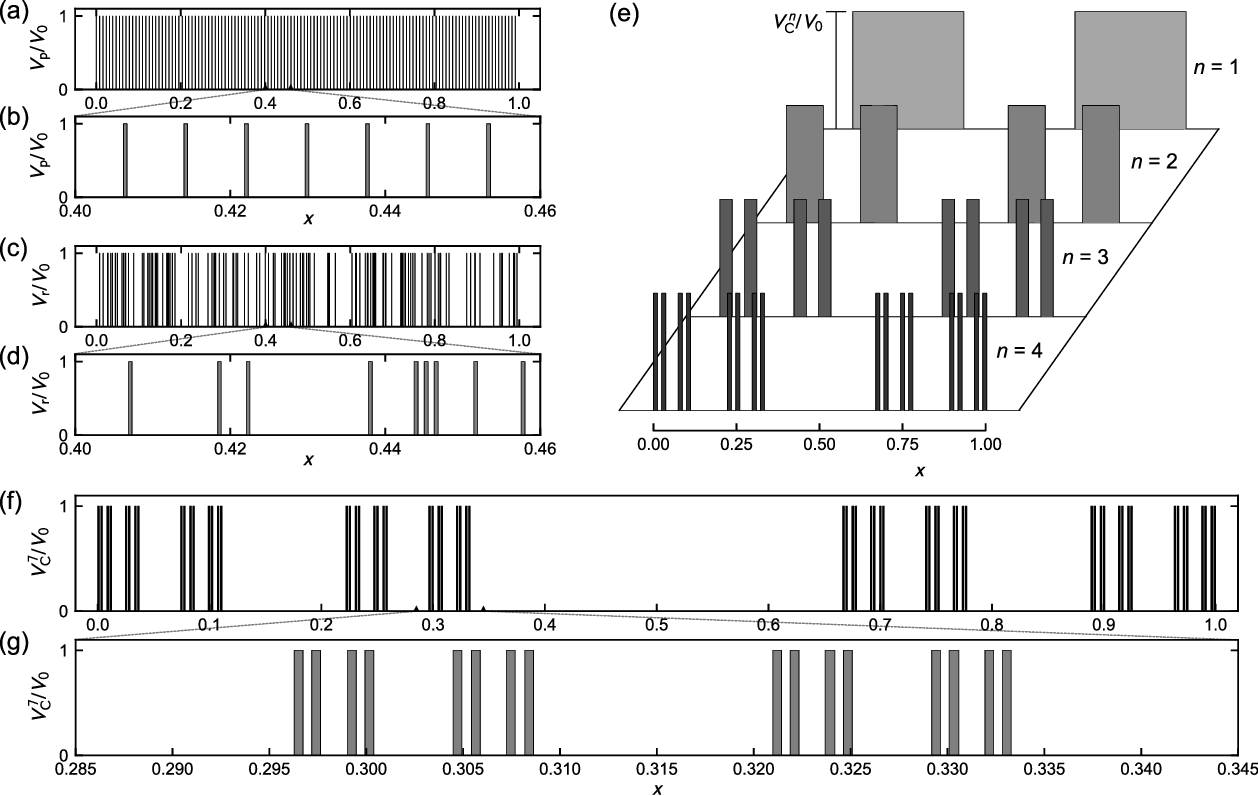}%
\caption{\label{fig:potentials} Spatial profiles of the potentials investigated in this study. 
(a) Periodic potential $V_\mathrm{p}(x)$ composed of equally spaced barriers.
The barrier count ($N_\mathrm{b}=128$) and width are matched to those of the seventh-generation Cantor potential to ensure a direct comparison. 
(b) Magnified view of a segment of the periodic potential.
(c) Random potential $V_\mathrm{r}(x)$, where barrier positions are distributed uniformly at random.
Although the limited plotting resolution may visually suggest overlapping barriers, the numerical model strictly imposes a non-overlapping constraint.
(d) Magnified view of the random potential.
(e) Iterative construction of Cantor potential $V_\mathrm{C}^n(x)$ for generations $n=1$--4.
(f) The seventh-generation Cantor potential $V_\mathrm{C}^7(x)$ used in the main analysis. 
(g) Magnified view of $V_\mathrm{C}^7(x)$, highlighting the self-similar fractal structure.
}
\end{center}
\end{figure*}

Figure~\ref{fig:potentials} illustrates the spatial profiles of the potential landscapes considered in this work.
The periodic potential $V_\mathrm{p}(x)$, shown in Fig.~\ref{fig:potentials}(a), serves as a reference for ordered systems.
It consists of a regular array of identical barriers.
As detailed in Sec.~\ref{sec:method}, the barrier parameters (height, width, and total count $N_\mathrm{b}=128$) are identical to those of the seventh-generation Cantor potential, ensuring that any spectral differences arise solely from the spatial arrangement.
The regularity of the lattice is clearly visible in the enlarged view presented in Fig.~\ref{fig:potentials}(b).

The random potential $V_\mathrm{r}(x)$ is depicted in Fig.~\ref{fig:potentials}(c).
Here, the barrier positions are determined by independent uniform random variables.
To ensure reproducibility, the random sequence was generated using standard pseudo-random number generators with a fixed seed.
As shown in the magnified view [Fig.~\ref{fig:potentials}(d)], the potential any translational symmetry.
Note that while the density of barriers may cause them to appear indistinguishable or overlapping in the full-scale plot, they are strictly non-overlapping in the actual numerical calculation.

The hierarchical structure of the Cantor potential is displayed in Figs.~\ref{fig:potentials}(e)--(g).
Figure~\ref{fig:potentials}(e) demonstrates the iterative generation process of the Cantor set for $n=1$ to $4$.
The fully developed seventh-generation potential $V_\mathrm{C}^7(x)$, which is the primary focus of our spectral analysis, is shown in Fig.~\ref{fig:potentials}(f).
The magnified view in Fig.~\ref{fig:potentials}(g) reveals the characteristic self-similar geometry, where the pattern of gaps and barriers reproduces itself across different length scales.

\subsection{Density of states}

\begin{figure}
\begin{center}
\includegraphics[width=8.5cm]{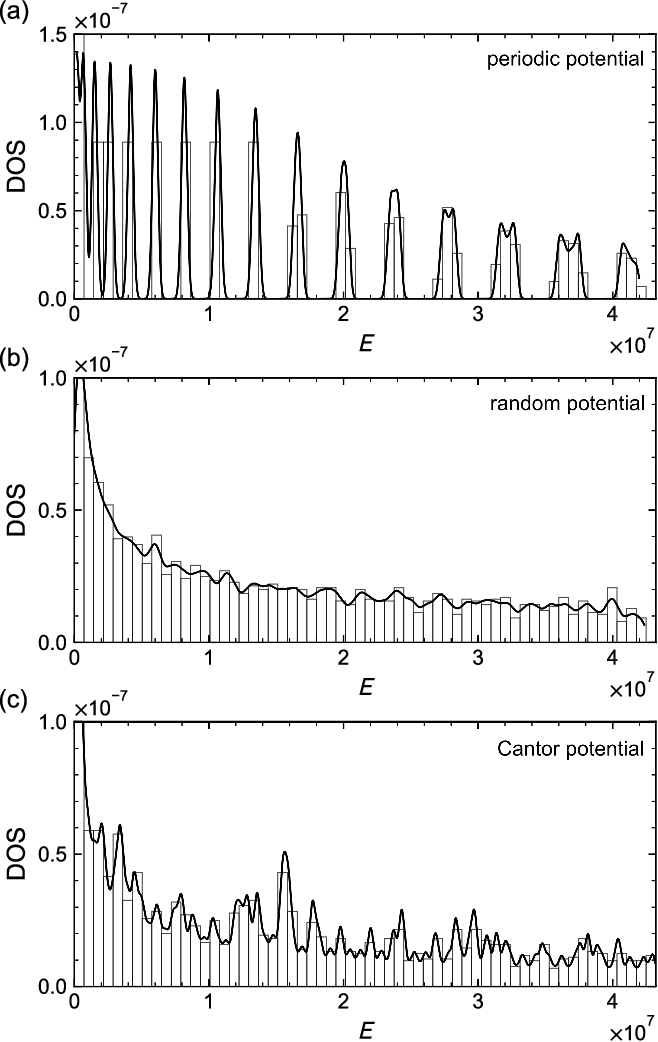}%
\caption{\label{fig:DOS_hist} Density of states (DOS) for (a) the periodic, (b) the random, and (c) the Cantor potentials.
The DOS is presented as a normalized histogram (with 60 bins) such that the total area integrates to unity.
Solid curves represent smoothed distributions obtained via kernel density estimation (KDE).
Note that the pronounced DOS peaks in the lowest-energy bins for the random and Cantor potentials exceed the vertical axis range and are truncated; the corresponding peak values are approximately $1.53\times 10^{-7}$ and $1.70\times 10^{-7}$, respectively.}
\end{center}
\end{figure}

We begin our analysis of the spectral properties by examining the density of states (DOS) for the three potential landscapes.
Figure~\ref{fig:DOS_hist} displays the DOS histograms, constructed by binning the eigenvalues over the entire computed energy range.
Each histogram is normalized to unity to facilitate a direct comparison of the spectral weights.
To mitigate artifacts arising from the finite bin width and to visualize the underlying distribution, we also plot smoothed curves obtained by kernel density estimation (KDE) using a Gaussian kernel.
The bandwidth of the KDE was optimized to capture the global spectral features, and we confirmed that the resulting profiles are robust against moderate variations of this parameter.

For the periodic potential [Fig. \ref{fig:DOS_hist}(a)], a distinct band structure is observed, in accordance with Bloch's theorem.
Although the finite bin resolution limits the visibility of fine details within each band (spanning only one or a few bins), the global trend is evident: the band widths increase with energy.
This broadening reflects the increase in kinetic energy relative to the potential barrier height, which enhances tunneling probability between adjacent wells and consequently widens the allowed energy bands.
This interpretation is further supported by our check that increasing the barrier height $V_0$ leads to narrower bands and wider gaps.

The DOS for the random potential is shown in Fig. \ref{fig:DOS_hist}(b).
A salient feature is the exceptionally large spectral density in the lowest-energy bin (value $\approx 1.53\times 10^{-7}$), which exceeds the vertical scale of the plot.
This peak corresponds to states localized in the widest potential cavities (Lifshitz tails)~\cite{Lifshitz1964, Lifshitz1988}.
Apart from this low-energy accumulation, the DOS exhibits a continuous and smooth distribution devoid of any discernible band gaps, consistent with the complete absence of translational symmetry and the disordered nature of the system.

In contrast, the Cantor potential [Fig. \ref{fig:DOS_hist}(c)] exhibits a highly intricate DOS reflecting its underlying fractal geometry.
While the global shape of the DOS is relatively stable against changes in system size and potential height, fine spiky structures persist across energy scales.
This behavior indicates that the spectrum preserves the hierarchical fragmentation of energy bands arising from the self-similar potential modulation even after spatial discretization, clearly distinguishing the fractal system from both periodic and random limits.

\subsection{Energy dependence of level spacings}

\begin{figure}
\begin{center}
\includegraphics[width=8.5cm]{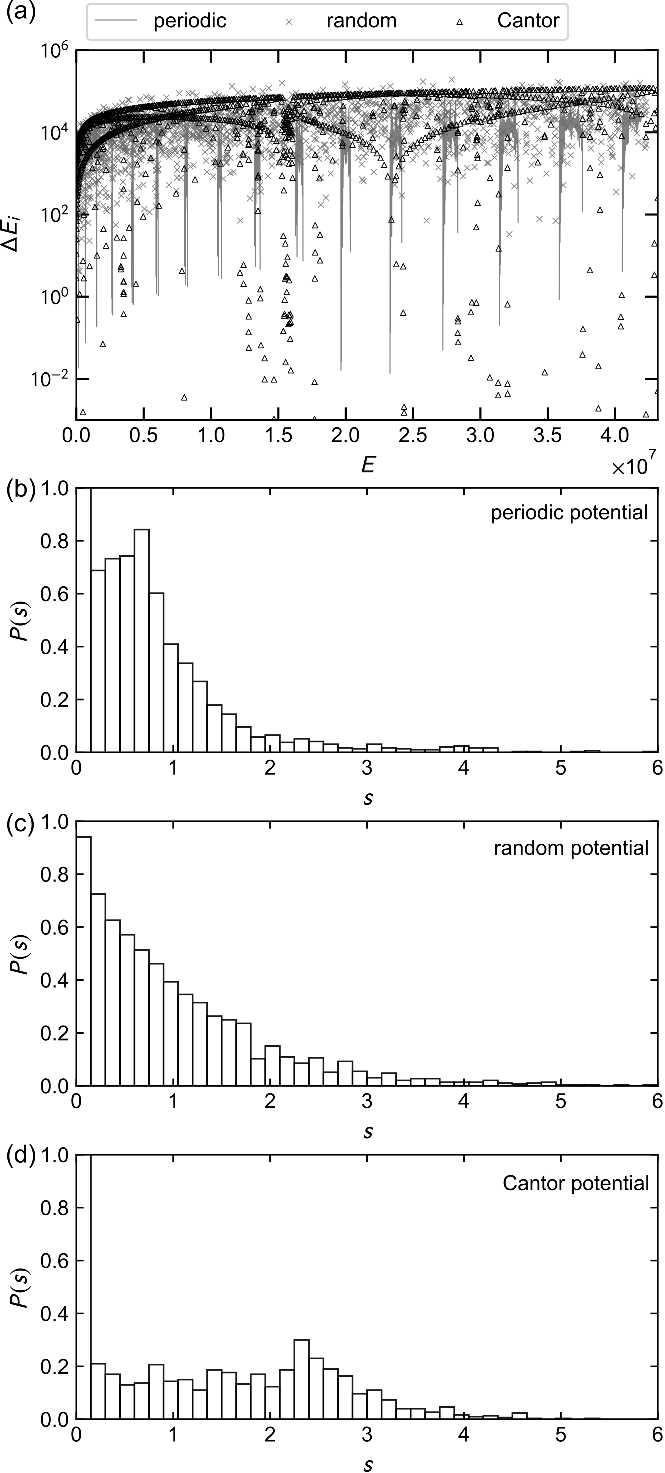}%
\caption{\label{fig:gap_analysis} (a) Energy dependence of the nearest-neighbor level spacings $\Delta E_i$.
Spacings smaller than $10^{-3}$ are regarded as numerically degenerate and excluded from the plot.
(b)--(d) Histograms of the normalized level spacings $s_i$ for the periodic, random, and Cantor potentials, respectively.
All distributions are normalized so that the total area equals unity.}
\end{center}
\end{figure}

In Fig.~\ref{fig:gap_analysis}(a), we present the energy dependence of the nearest-neighbor level spacings, $\Delta E_i$, plotted against the mean energy of the adjacent levels.
Numerical gaps smaller than $10^{-3}$ are attributed to quasi-degeneracies and are excluded from this analysis.

For the periodic potential, the spectrum exhibits a clear separation into allowed bands and forbidden gaps.
The allowed bands are characterized by small level spacings ($10^{-1}$ to $10^5$), while the band gaps are significantly larger, on the order of $10^6$.
Within each allowed band, the level spacing is smallest near the band edges and increases toward the band center.
This U-shaped behavior corresponds to the divergence of the density of states at the band edges, a manifestation of the van Hove singularities characteristic of one-dimensional periodic systems~\cite{Aschcroft1976}.

In sharp contrast, the random potential displays a broad, cloud-like distribution of level spacings ranging from $10^2$ to $10^5$.
No distinct band structures or large gaps are discernible.
This continuous distribution reflects the lack of translational symmetry and the absence of resonant transmission channels that would otherwise form extended Bloch states.

The Cantor potential shows a markedly different spectral structure.
Instead of forming simple bands, the energy-gap plot exhibits characteristic filamentary structures extending across a wide energy range.
These structures, which correspond to level spacings of distinct scales, appear as elongated branches that occasionally intertwine or exhibit fragmentation.
This complex pattern originates from the self-similar nature of the potential, where gaps of different generations evolve and interact across energy scales.
In addition to these primary features, we observe scattered data points with small spacings that deviate from the main filaments.
Our wave-function analysis reveals that these are not numerical artifacts but originate from tunneling splittings associated with the global mirror symmetry of the potential.
Specifically, these quasi-degenerate pairs correspond to symmetric and antisymmetric superpositions of states localized in spatially separated, symmetric sub-regions.
Such features are not unique to the seventh generation but are already observed as developing branches in lower-generation Cantor potentials.

\subsection{Normalized gap analysis}
To characterize the local spectral correlations intrinsic to each potential, we analyze the probability distribution $P(s)$ of the unfolded level spacings $s_i$.
The unfolding procedure removes the global energy dependence of the density of states, allowing us to focus on the fluctuations arising from the potential geometry.
The resulting distributions for the three potential types are presented in Figs.~\ref{fig:gap_analysis}(b)--(d).

For the periodic potential [Fig. \ref{fig:gap_analysis}(b)], the distribution $P(s)$ shows a pronounced peak centered around $s \simeq 0.675$ and decays smoothly toward larger values.
The concentration of probability weight below $s=1$ indicates a predominance of closely spaced levels, which is a consequence of the continuous band structure.
The absence of additional structure suggests a relatively uniform degree of level repulsion within the allowed bands.

For the random potential [Fig.~\ref{fig:gap_analysis}(c)], the distribution profile changes markedly.
The distinct peak at $s \simeq 0.675$ observed in the periodic system disappears, and the distribution exhibits a smooth decay.
Compared to the periodic case, there is a noticeable enhancement of the probability density in the intermediate region around $s \approx 1.5$.
This redistribution of spectral weight reflects the breakdown of the regular band structure and the modification of level correlations due to disorder.

In contrast, the Cantor potential exhibits a qualitatively distinct behavior [Fig.~\ref{fig:gap_analysis}(d)].
The distribution $P(s)$ is highly non-trivial, displaying a clear bimodal structure with distinct peaks near $s \approx 0$ and $s \approx 2.3$.
This alternating pattern of dense and sparse spectral regions is a direct manifestation of the self-similar geometry, where the hierarchy of potential barriers generates gaps on multiple energy scales.

These results demonstrate that the local statistics of energy gaps provide a sensitive probe of underlying potential structure.
The Cantor potential, in particular, possesses a unique spectral signature---simultaneous level clustering and large gap formation---that is absent in both standard periodic and disordered systems.
We have confirmed that these distributional features are robust against reasonable variations in the smoothing parameter used for unfolding.

\subsection{Multifractality and inverse participation ratio}
\subsubsection{Multifractal dimension}

\begin{figure}
\begin{center}
\includegraphics[width=8.5cm]{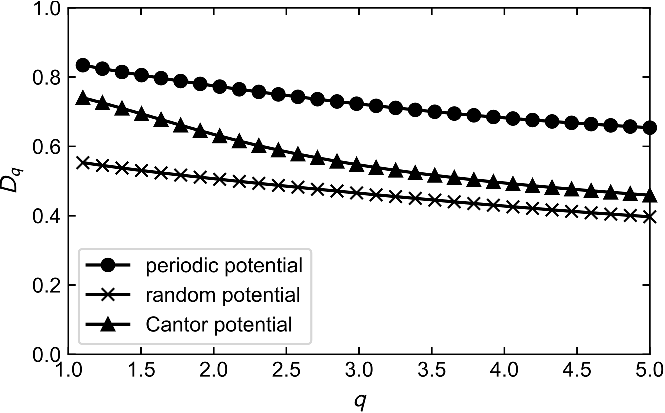}%
\caption{\label{fig:IPR_multi} 
Dependence of the generalized fractal dimension $D_q$ on the moment order $q$ for the periodic, random, and Cantor potentials.
The parameter $q$ ranges from $1.1$ to $5.0$.
Solid lines are guides to the eye.
}
\end{center}
\end{figure}

Figure~\ref{fig:IPR_multi} presents the generalized multifractal dimension $D_q$ as a function of the moment index $q$.
The parameter $q$, varied here from $1.1$ to $5.0$, acts as a filter for the wave-function amplitude: as $q$ increases, the generalized moment $P_q$ becomes increasingly dominated by region of high probability density $|\psi(x)|^2$.

For the periodic potential, $D_q$ exhibits a gradual decrease from approximately $0.8$ to $0.7$.
While theoretically $D_q=1$ for perfectly uniform plane waves, the observed deviation reflects the internal modulation of the Bloch states, which are concentrated within the potential wells.
Nevertheless, the values remain higher than those of the random potential and show the weakest dependence on $q$ among the three cases, characterizing the extended nature of the states.

In the case of the random potential, $D_q$ also decreases monotonically with $q$, but its overall magnitude is smaller than that of the periodic case.
This indicates enhanced localization induced by the random arrangement of potential barriers.

In contrast, the Cantor potential displays a distinct, intermediate behavior.
The values of $D_q$ fall strictly between the periodic and random limits.
Notably, a distinct decrease in dimension is observed around $q \simeq 2$.
Recalling that $D_2$ is determined by the IPR, this behavior signifies that the eigenstates are critical---neither fully extended like Bloch waves nor exponentially localized like Anderson states.

\subsubsection{Inverse participation ratio}

\begin{figure}
\begin{center}
\includegraphics[width=8.5cm]{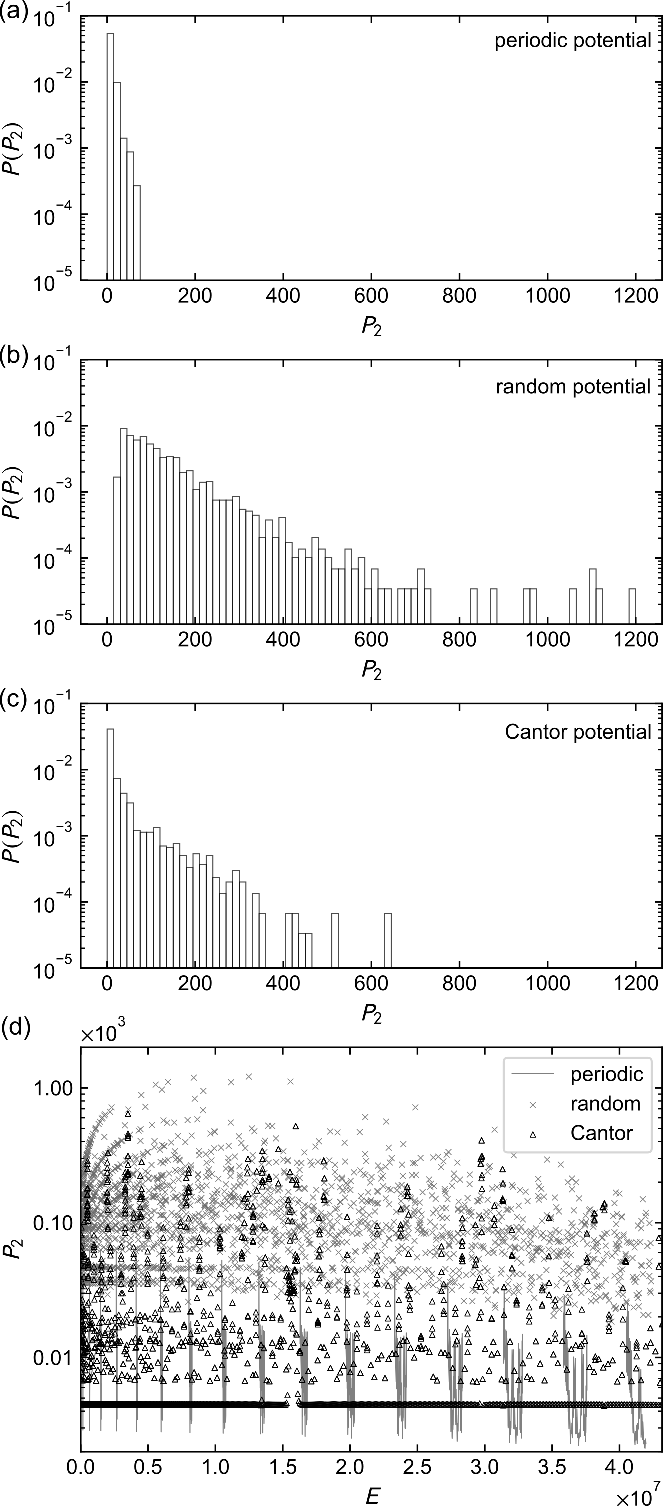}%
\caption{\label{fig:IPR} 
(a)--(c) Histograms of the inverse participation ratio (IPR, $P_2$) calculated for the periodic, random, and Cantor potentials, respectively.
The vertical axis represents the probability density $P(P_2)$ on a logarithmic scale, while the horizontal axis shows $P_2$ on a linear scale.
The histograms are constructed using 80 bins over the range $0 \leq P_2 \leq 1200$.
(d) Energy dependence of the IPR for the three potentials, plotted on a semi-logarithmic scale.
}
\end{center}
\end{figure}

To quantify the localization properties of the wave functions, we investigate the IPR, denoted as $P_2$.
Figures~\ref{fig:IPR}(a)--(c) display the histograms of the IPR values for the three potential landscapes.
In these plots, the vertical axis corresponds to the probability density $P(P_2)$, shown on a logarithmic scale to capture the broad distribution of values. 

For the periodic potential [Fig.~\ref{fig:IPR}(a)], the distribution is extremely narrow and concentrated at small $P_2$ values.
This indicates that the majority of eigenstates are spatially extended, consistent with the Bloch theorem which predicts delocalized states extending over the entire lattice.

In sharp contrast, the random potential [Fig.~\ref{fig:IPR}(b)] exhibits a broad distribution shifted significantly toward larger $P_2$ values.
States with small $P_2$ are virtually absent, implying that all eigenstates are strongly localized.
This result is consistent with Anderson localization in one-dimensional disordered systems, where all states are expected to be exponentially localized~\cite{Anderson1958, Kramer1993}.

The Cantor potential [Fig.~\ref{fig:IPR}(c)] presents an intermediate and distinct character.
The distribution is broad and covers a wide range of $P_2$ values, indicating the coexistence of relatively extended (low $P_2$) and highly localized (high $P_2$) states.
This broadness reflects multiscale nature of the potential, which supports wave functions with varying degrees of spatial confinement corresponding to the hierarchical gap structure.

The energy dependence of the IPR, shown in Fig.~\ref{fig:IPR}(d), offers further insight into the spectral properties.
For the periodic potential, the IPR remains uniformly small across the energy spectrum, as the eigenstates extend over multiple potential barriers within the allowed bands.

In the random potential, the IPR remains large values over the entire energy range.
No mobility edge or energy region corresponding to extended states is observed, confirming that disorder induces strong localization for all eigenstates in this one-dimensional system.

For the Cantor potential, the IPR fluctuates significantly, spanning both low and high values over a wide energy range.
Notably, the states with the minimum observed $P_2$ value ($\approx 4.5$) can be identified as states confined within the central (and largest) potential well formed by the first-generation construction.
The spatial extent of these states is essentially determined by the width of this central cavity ($L/3$), resulting in a nearly constant, relatively low $P_2$ that is insensitive to energy variations.
On the other hand, the finer potential structures associated with higher-generation barriers induce stronger confinement, leading to states with much larger $P_2$.
This energy-dependent mixture of localization lengths---neither purely extended nor simply localized---is a hallmark of critical states and is consistent with the multifractal characteristics discussed in the preceding section.

\subsection{Scaling analysis of the integrated density of states}

\begin{figure}
\begin{center}
\includegraphics[width=8.5cm]{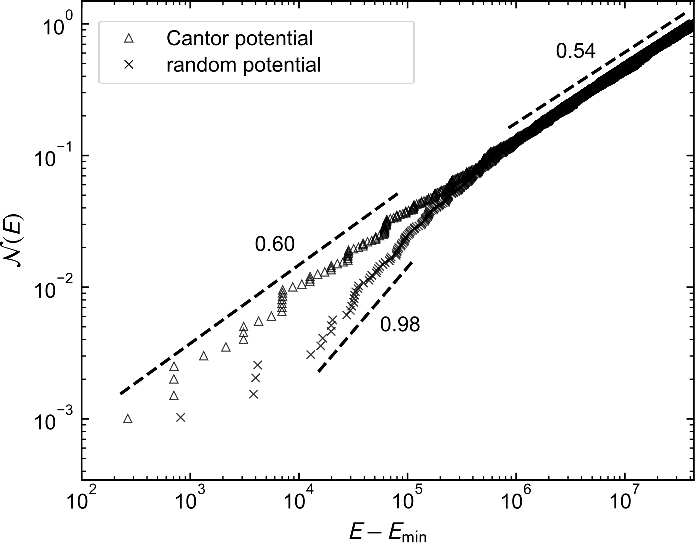}%
\caption{\label{fig:IDS_scaling} 
Scaling behavior of the integrated density of states (IDS), $\mathcal{N}(E)$, for the Cantor and random potentials.
The IDS is plotted against the excitation energy $E-E_\mathrm{min}$ on a log-log scale.
Dashed lines indicate power-law fits.
For the random potential, the fit is restricted to $E-E_\mathrm{min} \gtrsim 10^4$ to exclude the fluctuation-dominated region at the lower band edge.
In the low-energy regime, the Cantor potential exhibits a distinct power-law scaling exponent ($\alpha \simeq 0.60$) compared to the random potential ($\alpha \simeq 0.98$).
In the high-energy regime, the scaling exponents for both potentials converge to approximately $0.5$, indicating the dominance of the kinetic energy term (free-particle-like behavior) over the potential details.}
\end{center}
\end{figure}

Figure~\ref{fig:IDS_scaling} presents the scaling analysis of the integrated density of states (IDS), contrasting the Cantor and random potentials.
In the low-energy regime ($10^2 \lesssim E -E_\mathrm{min} \lesssim 10^5$), a marked difference distinguishes the two systems.
The IDS follows a power-law scaling form $\mathcal{N}(E) \sim (E- E_\mathrm{min})^\alpha$.

For the Cantor potential, the scaling exponent in this low-energy region is estimated to be $\alpha \simeq 0.60$.
This deviation from unity ($\alpha = 1$) reflects the anomalous spectral dimension induced by the fractal potential structure.

For the random potential, the behavior is qualitatively different.
We note that the data points in the lowest energy region ($E - E_\mathrm{min} \lesssim 10^4$) correspond to rare states localized in exceptionally wide potential cavities formed by fluctuation (Lifshitz tails) and are subject to significant statistical fluctuations in a finite system; hence, they were excluded from the fitting analysis.
In the region immediately above this threshold, the random potential exhibits a scaling exponent of $\alpha \simeq 0.98$.
This value is distinctly larger than that of the Cantor potential ($\alpha \simeq 0.60$), highlighting that the spectral properties of the Cantor system are fundamentally different from those of a generic disordered system.

In contrast, in the high-energy regime ($E - E_\mathrm{min} \gtrsim 10^6$), the scaling behaviors of the Cantor and random potentials converge.
Both systems exhibit exponents approaching $\alpha \simeq 0.54$.
This value is remarkably close to the theoretical exponent for a one-dimensional free particle ($\alpha = 0.5$), where $\mathcal{N}(E) \sim \sqrt{E}$.
This convergence suggests that, at sufficiently high energies, the kinetic energy term dominates the Hamiltonian.
In this short-wavelength limit, the wave functions oscillate rapidly enough to average out the detailed spatial variations of the potential, recovering the free-particle density of states.

We confirmed that the qualitative features of these scaling behaviors are robust against variations in the potential height.
Furthermore, although the present data correspond to the seventh-generation Cantor potential, we observed that the scaling behavior in the low-energy region has already converged by the sixth generation.
This implies that the low-energy spectral properties are determined by the coarser, large-scale structures of the potential (lower generations), while the finer structures introduced at higher generations primarily modify the high-energy spectrum.

These results demonstrate that the Cantor potential possesses a genuinely fractal energy spectrum, characterized by a non-trivial scaling exponent $\alpha$ in the low-energy regime that differs from both the disordered (random) and free-particle limits.

\section{Discussion\label{sec:discussion}}
\subsection{Hierarchical gap statistics in the Cantor potential}

The distinctive filamentary structures observed in the level-spacing statistics [Fig.~\ref{fig:gap_analysis}(a)] originate from the discrete nature of the confinement lengths inherent to the Cantor potential.
Since the potential consists of a hierarchy of wells (cavities) with varying widths ($L/3, L/9, \dots$), eigenstates confined within wells of the same generation possess nearly identical energies and level spacings, while their absolute energies spread into narrow bands due to tunnel coupling between the wells.
Consequently, the nearest-neighbor gaps remain approximately constant over finite energy intervals, forming the observed horizontal branches.
This behavior is a direct manifestation of the self-similar geometry, where a large number of potential segments share identical characteristic length scales, leading to macroscopic quasi-degeneracies in the spectrum.

However, a closer inspection reveals that this ideal hierarchical structure is partially disrupted in certain energy windows (around $E \approx 1.6\times 10^6$).
In these regions, the eigenstates are predominantly localized within the narrowest potential wells corresponding to the highest generation of the Cantor set.
Due to the vast number of such minimal-width wells, the degeneracy is lifted by finite tunneling couplings and boundary effects, resulting in a dense and complex distribution of energy levels.
This leads to a proliferation of possible spatial configurations for the wave functions, thereby inducing spectral irregularities that blur the sharp filamentary features.
Essentially, the finest level of the potential structure acts as a source of local disorder, introducing complexity into the gap statistics.

At higher energies, the wave functions become less sensitive to the finest potential details and extend over wider spatial regions.
In this regime, the spectral properties are governed by the coarser potential structures (lower generations, e.g., $n=6$), and the clear, hierarchical clustering of gap sizes is recovered.
Overall, the gap statistics of the Cantor potential exhibit a dual character: globally, they encode the deterministic, self-similar geometry of the potential, while locally they display nontrivial fluctuations driven by the interplay between the multitude of finest-scale wells.
These findings confirm that the fractal geometry is not merely a spatial feature but is deeply imprinted onto the energy spectrum itself.

\subsection{Multifractality and localization nature}

The $q$-dependence of the generalized fractal dimension $D_q$ reveals fundamental differences in the localization properties of the wave functions among the three potential landscapes.
Strictly speaking, $D_q$ is defined through a finite-size scaling analysis, measuring how the moments of the wave function scale with the system size $L$.
In the present study, however, the system size is fixed, and the analysis represents a snapshot at a specific spatial resolution.
Nevertheless, this effective $D_q$ serves as a robust quantitative indicator for comparing the degree of spatial confinement and heterogeneity across different potentials.

A distinguishing feature of the Cantor potential is the pronounced decrease of $D_q$ around $q\simeq 2$.
Since large positive $q$ values weigh the regions of high probability density amplitude, this behavior implies a strong spatial intermittency, where sharp peaks and sparse regions coexist within the wave functions.
Such a non-trivial $q$-dependence contrasts sharply with the nearly constant behavior observed in the periodic potential (extended states) and the uniformly suppressed values in the random potential (localized states).
This confirms that the Cantor potential supports ``critical'' wave functions that exhibit multifractal properties, distinct from both standard extended and exponentially localized states.

The statistics of the IPR provide further insight into these spectral characteristics.
For the periodic potential, the uniformly small $P_2$ reflect the extended nature of Bloch states.
Conversely, the random potential exhibits consistently large $P_2$, a result that is fully consistent with the theory of Anderson localization, which dictates that all eigenstates in one-dimensional disordered systems are exponentially localized even for weak disorder~\cite{Anderson1958, Kramer1993}.

The Cantor potential, owing to its fractal structure, displays a hybrid behavior characterized by a broad distribution of IPR values.
This signifies the coexistence of states with widely varying localization lengths---ranging from quasi-extended states to those strongly confined within narrow potential cavities.
This diversity is a hallmark of critical systems, where the wave functions are neither simple plane waves nor simple localized peaks but possess a self-similar structure intermediate between the two limits.

Finally, we note the energy dependence of the localization properties in the high-energy regime.
It is observed that the IPR does not exhibit a sharp cutoff or a sudden drop to zero.
Instead, the distribution of highly localized states (high $P_2$) merely becomes sparse as the energy increases.
While states with relatively large $P_2$ values persist to some extent even at high energies, their occurrence becomes noticeably less frequent compared to the low-energy region.
This gradual thinning of the high-IPR population suggests that, although the short-wavelength kinetic effects begin to dominate, the potential's fractal structure continues to support occasional localized states, albeit with diminishing probability.

\subsection{Spectral dimensionality and Weyl asymptotics}

The scaling analysis of the IDS, $\mathcal{N}(E)$, elucidates the fundamental difference between the disordered and fractal spectra.
For the random potential, the IDS in the low-energy regime follows a power-law behavior with an exponent of $\alpha \simeq 0.98$.
This near-unity exponent implies that, on average, the localized states emerge with a constant density as energy increases, resulting in a linear growth of $\mathcal{N}(E)$.
Physically, this suggests that while the eigenstates are localized, the randomness creates a spatially uniform distribution of potential fluctuations, leading to a gapless spectrum with a finite average density of states.

In stark contrast, the low-energy scaling exponent of the Cantor potential significantly deviates from unity.
This deviation is a direct manifestation of the fractal geometry in the quantum energy spectrum.
Specifically, the spectrum is characterized by a ``spectral dimension'' that differs from the physical dimension of the system.
Remarkably, the obtained exponent ($\alpha \simeq 0.60$) is close to the Hausdorff dimension of the Cantor set, $\log 2 / \log 3 \approx 0.631$.
This proximity suggests that the geometry fractality of the potential is deeply imprinted onto the scaling properties of the quantum spectrum.
It demonstrates that for a deterministic fractal potential, the distribution of eigenvalues is governed by the underlying self-similar geometry, providing clear evidence of critical spectral features.

In the high-energy regime, the scaling exponents for both the Cantor and random potentials converge to approximately $0.54$.
This agreement is noteworthy, given the clear differences remaining in the detailed DOS.
The coincidence arises because the IDS, being an integral of the DOS, averages out local spectral fluctuations.
Moreover, at high energies, the de Broglie wavelength of the wave function becomes shorter than the characteristic width of the potential barriers, rendering the detailed potential structure effectively irrelevant.
In this semiclassical limit, the IDS follows a Weyl-type asymptotic behavior, $\mathcal{N}(E) \propto \sqrt{E}$ (i.e., $\alpha = 0.5$ for 1D free particles)~\cite{Weyl1911, Gutzwiller1990, Pastur1994}, confirming that microscopic details are averaged out in the short-wavelength limit~\cite{Brack2003}.

A closer inspection of the IDS for the Cantor potential reveals that, while $\mathcal{N}(E)$ increases monotonically, it does not do so smoothly.
Instead, the IDS exhibits a modulated growth pattern, characterized by alternating intervals of steep and shallow slopes.
The regions of steep slope correspond to the energy clusters (bands) where eigenstates are densely distributed, whereas the regions of shallow slope correspond to the pseudogaps where the density of states is significantly suppressed but not strictly zero.
This behavior is reminiscent of a ``Devil's staircase,'' yet distinct in the ``steps'' possess finite slopes rather than forming flat plateaus.
This feature originates from the self-similar fractal structure of the Cantor potential and reflects a multiscaling nature, where local scaling fluctuates depending on the energy window.

For higher-generation Cantor potentials, the hierarchy of spectral gaps becomes more intricate.
Investigating how the control of potential barrier heights affects this staircase structure and whether it leads to a smooth crossover in the scaling behavior remains an interesting subject for future work.

\section{Conclusion\label{sec:conclusion}}

In this paper, we numerically studied the gap statistics and fractal properties of quantum states in a one-dimensional Cantor potential.
We demonstrated that the Cantor potential exhibits gap statistics that are fundamentally distinct from those of conventional periodic and random potentials.
Specifically, the nearest-neighbor level spacings develop a hierarchical, filamentary structure in the energy--spacing plane.
The corresponding normalized gap distribution deviates from random-matrix predictions, indicating that the self-similar geometry of the Cantor potential is directly reflected in the spectral statistics.

A multifractal analysis of the eigenstates, supported by the IPR, revealed that the Cantor potential supports a coexistence of quasi-extended and localized states.
This results in critical quantum states that are neither fully delocalized like Bloch waves nor exponentially localized like Anderson states.
Such behavior highlights the nontrivial nature of wave-function confinement induced by deterministic fractal structures, which differs qualitatively from disorder-induced localization.

Furthermore, the scaling analysis of the IDS demonstrated that, in the low-energy regime, the geometric fractal nature of the Cantor potential is directly imprinted onto the quantum spectrum.
The observed power-law scaling, with an exponent distinct from both the random and free-particle limits, provides clear evidence that the hierarchical structure of the potential governs the spectral organization.
At higher energies, we observed a crossover to Weyl-type behavior, where short-wavelength kinetic effects wash out the fine fractal details.

Taken together, our results clarify how the self-similar structure of a deterministic fractal potential manifests itself in both the energy spectrum and the statistical properties of wave functions.
The deterministic control of spectral dimensionality and localization properties provided by the Cantor potential suggests a route toward geometry-driven wave-function engineering.
In contrast to random systems where localization is statistical and uncontrollable, the Cantor potential offers a deterministic route to generate critical states.
This feature is particularly promising for designing energy-selective filters or wave-guiding devices in photonics and acoustics, where specific spectral gaps and localization lengths are required.
More broadly, our findings shed light on quantum transport in fractal environments and establish the Cantor potential as a paradigmatic model for studying criticality and multifractality in low-dimensional quantum systems.

\end{document}